\documentclass[useAMS,usenatbib]{mn2e}

\usepackage{graphicx}

\title[On time dilation in quasar light curves]{On time dilation in
 quasar light curves}
\author[M. R. S. Hawkins]{M. R. S. Hawkins$^{1}$\thanks{E-mail:
 mrsh@roe.ac.uk}\\
$^{1}$Institute for Astronomy (IfA), University of Edinburgh,
 Royal Observatory, Blackford Hill, Edinburgh EH9 3HJ, UK}
\begin{document}

\date{Accepted 1988 December 15. Received 1988 December 14; in original
form 1988 October 11}

\pagerange{\pageref{firstpage}--\pageref{lastpage}} \pubyear{2002}

\maketitle

\label{firstpage}

\begin{abstract}
In this paper we set out to measure time dilation in quasar light curves.
In order to detect the effects of time dilation, sets of light curves
from two monitoring programmes are used to construct Fourier power
spectra covering timescales from 50 days to 28 years.  Data from
high and low redshift samples are compared to look for the changes
expected from time dilation.  The main result of the paper is that quasar
light curves do not show the effects of time dilation.  Several
explanations are discussed, including the possibility that time dilation
effects are exactly offset by an increase in timescale of variation
associated with black hole growth, or that the variations are caused by
microlensing in which case time dilation would not be expected.
\end{abstract}

\begin{keywords}
quasars: general -- cosmology: observations.
\end{keywords}

\section{Introduction}
\label{sec1}

Time dilation (the stretching of time by a factor of $(1+z)$) is a
fundamental property of an expanding universe.  Given the success of
the the currently accepted cosmological model, which certainly implies
expansion, it is perhaps surprising that more attention has not been
paid to making direct measures of time dilation.  This must surely be
due in part to the fact that measures of time dilation can tell little
or nothing about cosmological parameters within the framework of a Big
Bang universe, but only whether or not the Universe is expanding.  Also,
it turns out to be surprisingly hard to formulate a conclusive test for
time dilation.  What is needed is an event or fluctuation of known rest
frame duration which can be observed at sufficiently high redshift with
an accuracy which enables the predicted stretching by a factor of
$(1+z)$ to be observed.

The study of gamma-ray bursts has generated considerable interest in
time dilation.  The uncertainty in the intrinsic timescales of the
bursts has has made it difficult to measure time dilation directly, so
earlier papers (for example \cite{d98}) concentrated on
demonstrating consistency when time dilation was allowed for.  When
redshifts of individual bursts became available more definitive tests
were possible \cite{c01}, but correcting the raw data for selection
effects involving an inverse correlation between luminosiy and timescale
has made the results hard to interpret with confidence.  Rather than
seeing gamma-ray bursts as providing a convincing test of time dilation,
it is probably safer to say that if time dilation is a property of the
Universe, then observations of gamma-ray bursts are consistent with
this.

The light curves of distant supernovae provide a much more promising
test for time dilation.  Early work by \cite{g01} which involved
measuring the light curve widths of a sample of distant supernovae
covering a wide range of redshifts, provided convincing evidence for the
presence of cosmological time dilation.  More recently, \cite{f05} have
examined in detail the spectral evolution of a high redshift supernova,
and shown that it is not consistent with no time dilation at a very high
confidence level.

In this paper we address the question of whether time dilation is seen
in quasar light curves.  We use the light curves of over 800 quasars
monitored on timescales from 50 days to 28 years to construct Fourier
power spectra for high and low redshift samples, and compare their
Spectral Energy Distributions (SEDs) to look for the effects of time
dilation. 

\section{Observations}
\label{sec2}

The observations upon which this paper is based come from two main
sources. The first group are part of a long term monitoring programme
undertaken by the UK Schmidt Telescope at Siding Springs Observatory in
Australia from 1975 till 2002.  The survey area comprises the central
20 square degrees of of the ESO/SRC Field 287, centred on 21h 28m,
-45$^{\circ}$ (1950).  Some 300 plates were taken of this field in
several passbands and over time intervals varying from an hour to 28
years.  Part of the monitoring programme involved taking a sequence of
plates in the $B_{J}$ (IIIa-J/GG395) passband every year from 1975 to
2002, with the exception of 1976, and in most years at least 4 plates
were obtained.  In order to monitor time variability on shorter
timescales, a second sequence of 24 plates was taken from 1983 to 1985,
covering the 2 year period as uniformly as possible.

\begin{figure*}
\centering
\begin{picture} (0,280) (260,0)
\includegraphics[width=1.0\textwidth]{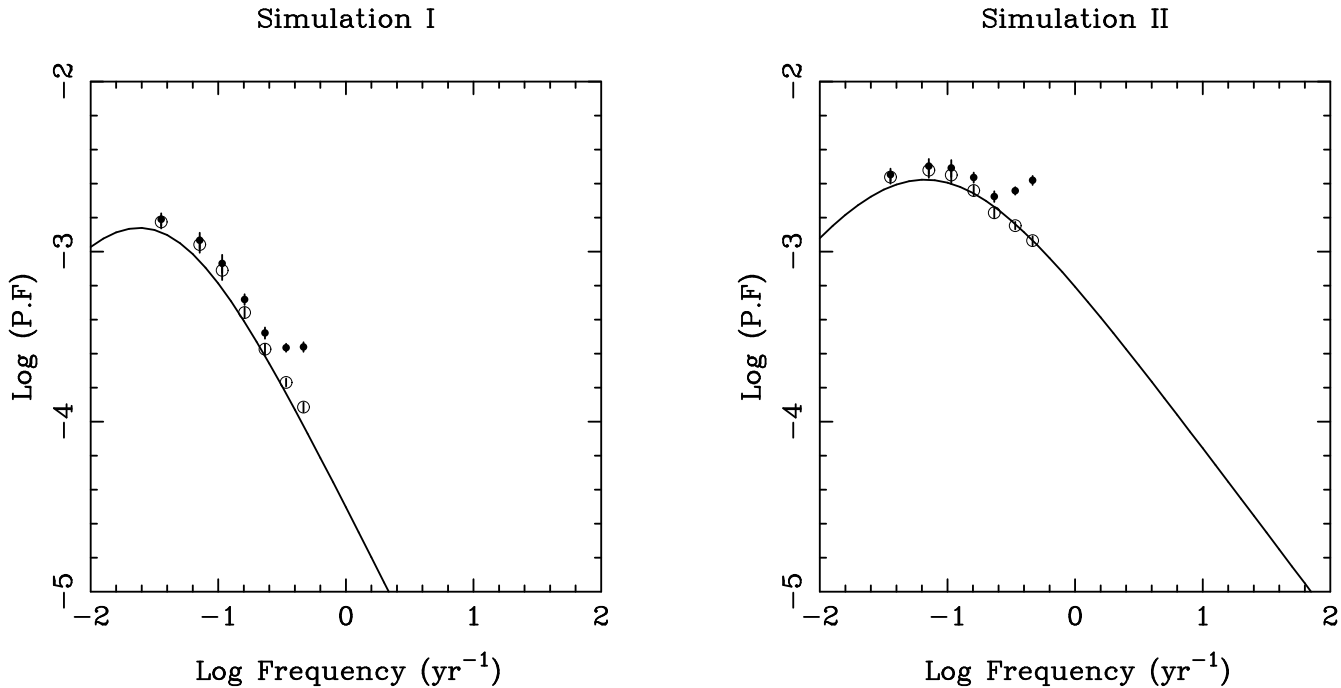}
\end{picture}
\caption{SEDs of sets of simulated light curves.  The solid lines show
the spectrum of variations used to construct the two different samples
of light curves.  Filled circles show the SEDs obtained by sampling as
for the real data, and the open circles show the result of applying the
sampling correction as described in the text.}
\label{fig0}
\end{figure*}
The plates fom the monitoring progamme were measured by the SuperCOSMOS
measuring machine at the Royal Observatory, Edinburgh to give catalogues
of flux, position and image quality parameters.  The raw magnitudes were
calibrated with deep CCD sequences, and the results paired up to give a
catalogue of some 200,000 objects to a completeness limit of
$B_{J} = 21.5$, with magnitudes for the epoch corresponding to each
plate.  The plates were reduced to the same zero-point using local
photometric transformations to minimise errors due to field effects
across the plates, and light curves were constructed for every object
in the catalogue from the mean magnitude for each year.  The overall
root mean square (rms) variation on the light curves due to photometric
errors was estimated by measuring the rms variation of the light curves
of samples of stars of similar colour and magnitude distribution to the
quasars.  The result, on the asumption that the stars were non-variable,
was an rms of $\pm 0.05$ mag.  

Throughout the period in which Field 287 was being observed, intensive
efforts were made to detect the quasars in the field.  Given the nature
of the survey material, quasars were most readily found from their
variability.  However, every other available technique was used to
supplement this method, including ultra-violet excess, objective prism
spectra, radio and X-ray surveys and two colour selection for high
redshift objects.  In the event most of the candidates which were
identified as quasars after spectroscopic follow up were found by more
than one technique, giving a good idea of the completeness of the final
quasar catalogue.  The total number of quasars so far confirmed with
redshifts now stands at over 1200 in the 20 square degrees of the survey
area.  Of these, 810 have no gaps in their light curves and make up a
complete sample with well defined magnitude and position limits.  This
sample was used for the light curve analysis. 

The second group of observations come from the monitoring programme of
the MACHO project \cite{a97}.  In this survey, which was primarily
designed to detect microlensing events in the Magellanic Clouds and
measure their light curves, an area containing several million stars
was observed on average every few days over a period of about eight
years.  In addition, there has been a determined effort to detect any
quasars lying in the survey area (\cite{g03}, \cite{d05} and references
therein).  Short timescale light curves for the 74 quasars with
adequate photometry were constructed by binning the observations at
50 day intervals, giving a total of 49 epochs for each quasar.  Of these,
68 lay within the luminosity bounds of the survey and were used in the
light curve analysis.

\section{Light curve analysis}
\label{sec3}

\subsection{Fourier power spectra}

\begin{figure*}
\centering
\begin{picture} (0,280) (260,0)
\includegraphics[width=1.0\textwidth]{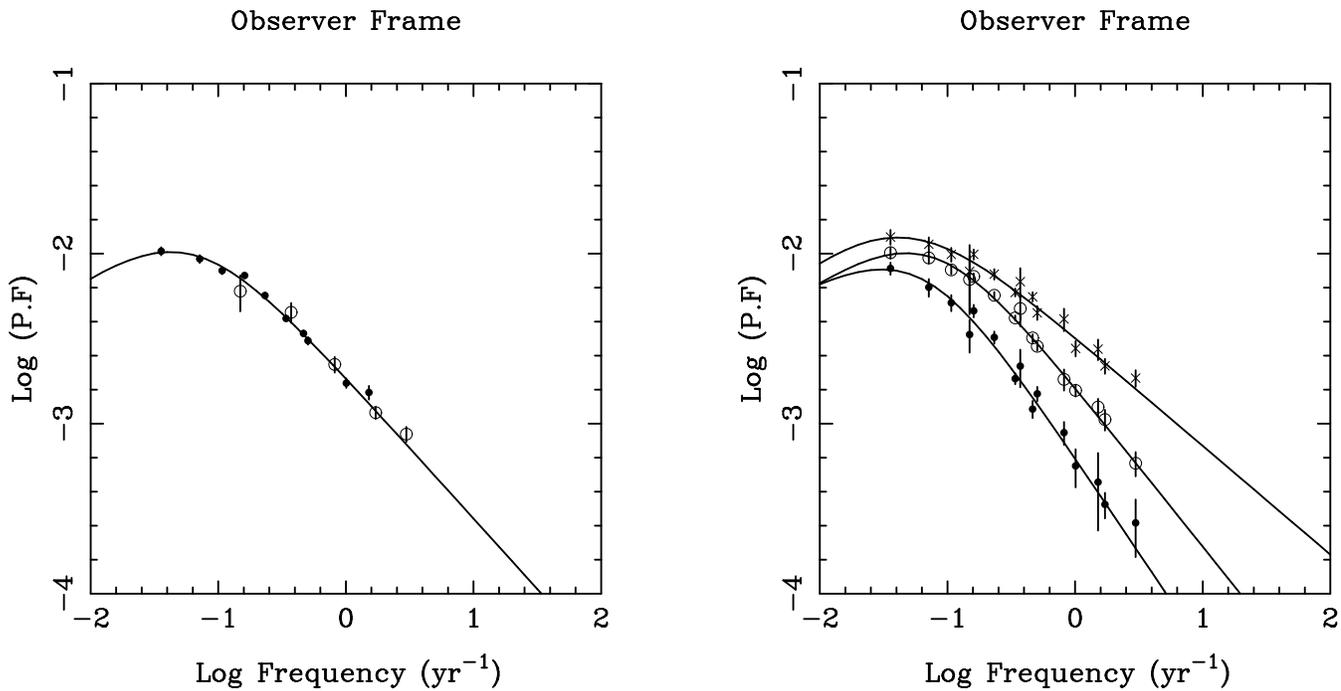}
\end{picture}
\caption{The left hand panel shows SEDs of light curves in the
 observer frame for quasars from the Field 287 survey (filled
 circles) and from the MACHO project (open circles).  The solid line is
 the best fit of the curve in Eq.~\ref{eqn2}.  The right hand panel 
 shows the same data divided into three magnitude ranges as defined in
 Table~\ref{tab1}.  In this case filled circles, open circles and
 stars represent high, medium and low luminosity bins respectively.
 Solid lines are fits to the data as for the left hand panel.}
\label{fig1}
\end{figure*}

The main purpose of this paper is to compare timescales of variation
in low and high redshift samples of quasars.  Fourier power spectrum
analysis will be used to quantify the variation on different timescales,
and to look for changes with redshift.  We define the Fourier power
spectrum $P(s)$ as:
 
\begin{eqnarray}
P(s_{i}) = \frac{\tau}{N} \left( \sum_{j=1,N} m(t_{j}) cos \frac{2 \pi
 i j} {N}\right)^{2} +
 \nonumber \\ \frac{\tau}{N} \left( \sum_{j=1,N}
 m(t_{j}) sin \frac{2 \pi i j} {N}\right)^{2}
\label{eqn1} 
\end{eqnarray}

\noindent
where $i$ runs over the $N$ equally spaced epochs of observation separated
by time $\tau$, and $m(t_{j})$ is the magnitude at epoch $t_{j}$.  In
the case of a sample of light curves, the integration for each frequency
continues over all sample members.

Since quasars typically vary on timescales of as little as a few months,
the power spectra of the light curves with several observations averaged
to give measures for yearly epochs must be corrected for the actual
sampling pattern.  This was done by constructing simulated light curves
with fully sampled epochs of observation, and with sampling identical
to the the times of the actual observations, with a resolution of 0.1
years.  The simulations were carried out along the lines advocated by
\cite{t95}.  Fourier power spectra were then calculated for sets of
1000 simulated light curves with full and actual sampling, and the
correction obtained as the difference between the two.  This correction
was then applied to the power spectrum of each of the observed light
curves.  The effect of the sampling correction is illustrated in
Fig.~\ref{fig0}, which shows SEDs for simulated light curves with widely
different modes of variation.  (For the purposes of this paper, we define
the SED as a plot of the product of Fourier power and frequency versus
frequency.)  The solid line shows the spectrum of variations on which
the simulations were based.  Also shown are the SEDs of the simulated
light curves sampled as for the actual data, and the effect of applying
the sampling correction. 

A second correction was also necessary to allow for the effects of
Poisson noise in the observed light curves as discussed by \cite{h07}.
The Poisson noise level was measured by calculating the Fourier power
spectra for large samples of stars, assumed to be non-variable.  This
correction, which was quite small, was subtracted from the integrated
power spectra of the samples of light curves.
\begin{figure*}
\centering
\begin{picture} (0,280) (260,0)
\includegraphics[width=1.0\textwidth]{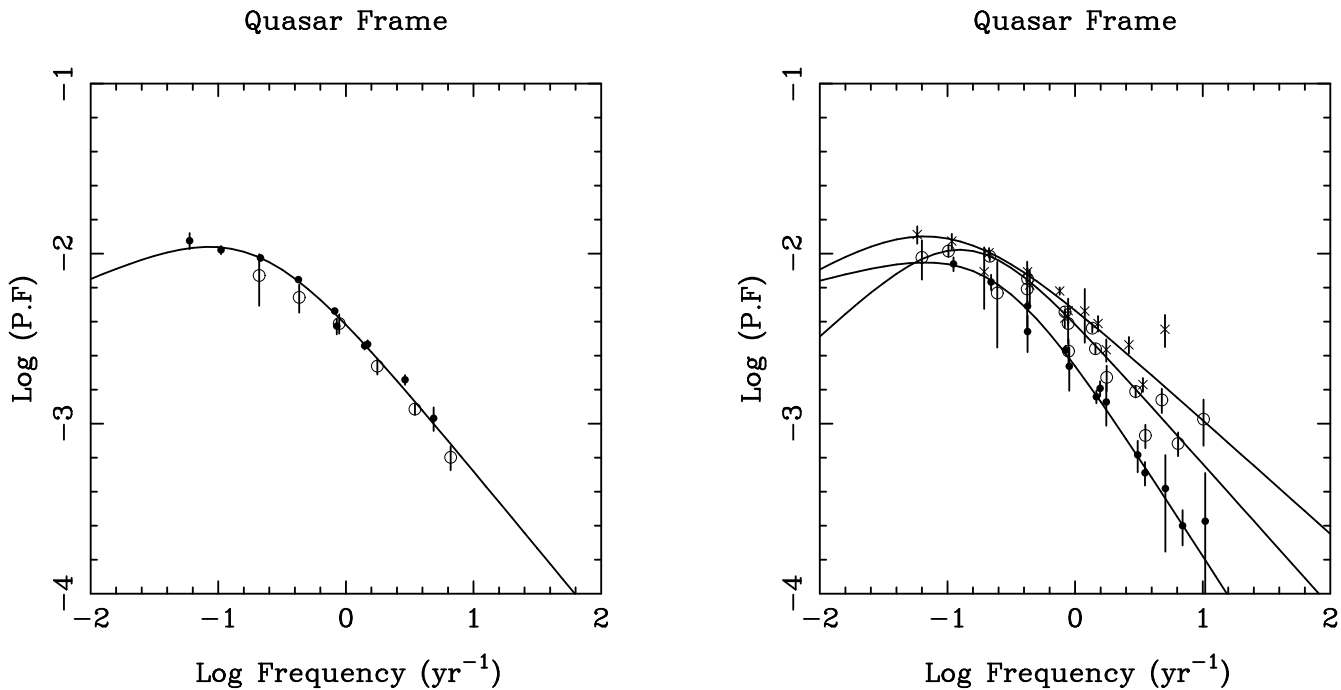}
\end{picture}
\caption{Same as for Fig.~\ref{fig1} but with a correction for time
 dilation applied to the SEDs.}
\label{fig2}
\end{figure*}
Fourier power spectra of the three sets of light curves defined in
Section~\ref{sec2} were calculated as described above, and are plotted
as SEDs in the left hand panel of Fig.~\ref{fig1}.  It will be seen that
all three SEDs are in good agreement in the areas of overlap, giving
confidence that there were no residual systematic effects in the data.
   
In order to obtain a measure of timescale and other parameters, the
SEDs were fitted by the function:
\begin{equation} 
P(f) = \frac{C}{\left(\frac{f}{f_{c}}\right)^{a}
 +\left(\frac{f}{f_{c}}\right)^{-b}}
\label{eqn2}
\end{equation}

\noindent
This function has the form of a power law rise from short timescales,
turning over at a characteristic timescale $ \tau = f_{c}^{-1}$.  The
maximum power density occurs at timescale $t_{max}$ given by:
\begin{equation}
t_{max} = \tau \left(\frac{a}{b}\right)^{\frac{1}{a-b}}
\label{eqn3}
\end{equation}

\noindent
The function $P(f)$ was fitted iteratively to the data in the left
hand panel of Fig.~\ref{fig1}, and the best fit is shown as a solid
line.  Best fit values for the parameters $t_{max}$ and $\tau$
are given in Table~\ref{tab1} together with the sample size, the
asymptotic value of the slope or power law index towards short timescales
($-a$), and the $\chi^2$ value of the fit.  With 14 degrees of freedom,
the fit is clearly satisfactory.

The errors on the model paramaters were estimated by repetitively
selecting half the sample at random, and carrying out the fitting
procedure.  The errors were then calculated from the measured
dispersion in the parameter values and are shown in Table~\ref{tab1}.

Preliminary examination of the left hand panel of Fig.~\ref{fig1}
suggests that the data follow a power law rise from high frequency
(short timescale) variation which breaks at a timescale of about 5
years to reach a maximum power density at around 25 years.  The
position of this maximum is however not well defined due to lack of
long timescale measurements.

\subsection{Magnitude effects}

The idea that there is a correlation between the way quasars vary and
their absolute magnitude or luminosity has a long history.  In
particular, several authors \cite {h94,c96,h00} have claimed to find
an anti-correlation between lumnosity and amplitude, in the sense that
for a sample of quasar light curves, more luminous quasars are seen to
vary over a smaller range of brightness than less luminous ones.  One
of the problems with this conclusion is that the observed amplitude is
clearly a function of the length of the run of observations, and so
can be confused with timescale of variability.  Fourier analysis
provides a way round this by giving measures of variability on different
timescales.  In the right hand panel of Fig.~\ref{fig1} the data in
the left hand panel is divided into three lumnosity ranges and the
SEDs fitted with the function $P(f)$ in Eq.~\ref{eqn2} as for the left
hand panel.

The three curves in the right hand panel of  Fig.~\ref{fig1} show
broadly the same features as the curve in the left hand panel, but it
is clear that the anti-correlation between luminosity and amplitude is
confirmed.  The maximum power density of the lowest luminosity quasars
is greater than the highest by a factor of 1.5.  In this case it
appears that the time span of the data is sufficient to resolve the
degeneracy between timescale and amplitude.

A striking feature of the three SEDs is the difference in the power
law indices at high frequencies.  It appears that there is a very
marked decrease in the amount of short timescale variation as quasars
become more luminous.  The analysis of this intriguing result is
beyond the scope of the present paper, but we note it here because of
its possible effect on the measurement of time dilation, and will
discuss it further in Section~\ref{sec4}.

Of great interest is the timescale at which the three SEDs reach their
maximum power density.  Unfortunately, lack of longer timescale data
makes the exact positions uncertain, and it would be premature to
identify any trend of this parameter with luminosity.  In fact, all
three measures of $t_{max}$ in Table~\ref{tab1} are consistent within
the errors with having the same value.  

\section{Time dilation}
\label{sec4}

\subsection{The quasar frame}

\begin{table*}
\caption{Timescale parameters for magnitude and redshift limited
samples of quasar light curves in the observer and quasar frames.}
\label{tab1}
\centering
\vspace{5mm}
\begin{tabular}{l c r r r c c r}
\hline\hline
 & & & & & & & \\
 sample & mean & size & $\tau$ (years) & $t_{max}$ (years)& index &
 $t_{ref}$ (years) & $\chi^{2}$ \\
 & & & & & & & \\
\hline
 & & & & & & & \\
\multicolumn{7}{c}{observer frame} \\
 & & & & & & & \\
\hline
 & & & & & & & \\
 all quasars & & 878 & 17.0 $\pm 1.6$ & 23.2 $\pm 0.9$ &
 -0.83 $\pm 0.02$ & 2.4 & 14.8 \\
 & & & & & & & \\
 $-23.5 < M_{B}$ & $<M_{B}> = -22.64$ & 217 & 23.2 $\pm 1.6$ &
 23.5 $\pm 2.1$ & -0.64 $\pm 0.02$ & 1.3 & 11.1 \\
 $-25.5 < M_{B} < -23.5 $  & $<M_{B}> = -24.55$ & 484 & 13.3 $\pm 0.9$ &
 20.6 $\pm 1.1$ & -0.94 $\pm 0.02$ & 2.5 & 6.1 \\
 \hspace{11.5mm} $M_{B} < -25.5$ & $<M_{B}> = -26.09$ & 177 &
 15.6 $\pm 1.4$ & 31.4 $\pm 6.2$ & -1.12 $\pm 0.05$ & 5.0 & 12.9 \\
 & & & & & & & \\
 $z < 1.0$ & $<z> = 0.765$ & 171 & 16.3 $\pm 1.9$ & 27.7 $\pm 5.0$ &
 -0.81 $\pm 0.05$ & 2.2 & 9.6 \\
 $z > 1.0$ & $<z> = 1.711$ & 469 & 14.2 $\pm 1.0$ & 20.3 $\pm 1.0$ &
 -0.88 $\pm 0.02$ & 2.2 & 7.3 \\
 & & & & & & & \\
\hline
 & & & & & & & \\
\multicolumn{7}{c}{quasar frame} \\
 & & & & & & & \\
\hline
 & & & & & & & \\
 all quasars & & 878 & 5.2 $\pm 0.4$ & 11.5 $\pm 0.6$ &
 -0.91 $\pm 0.03$ & 0.9 & 16.5 \\
 & & & & & & & \\
 $-23.5 < M_{B}$ & $<M_{B}> = -22.64$ & 217 & 10.7 $\pm 4.2$ &
 14.3 $\pm 7.6$ & -0.67 $\pm 0.05$ & 0.7 & 30.6 \\
 $-25.5 < M_{B} < -23.5 $  & $<M_{B}> = -24.55$ & 484 & 7.0 $\pm 0.4$ &
 7.8 $\pm 0.8$ & -0.84 $\pm 0.02$ & 1.0 & 37.4 \\
 \hspace{11.5mm} $M_{B} < -25.5$ & $<M_{B}> = -26.09$ & 177 &
 4.1 $\pm 0.3$ & 14.4 $\pm 3.4$ & -1.17 $\pm 0.04$ & 1.5 & 5.1 \\
 & & & & & & & \\
 $z < 1.0$ & $<z> = 0.765$ & 171 & 7.4 $\pm 0.6$ & 15.2 $\pm 2.9$ &
 -0.87 $\pm 0.05$ & 1.2 & 14.9 \\
 $z > 1.0$ & $<z> = 1.711$ & 469 & 6.8 $\pm 0.6$ & 7.4 $\pm 0.4$ &
 -0.82 $\pm 0.03$ & 0.8 & 5.2 \\
 & & & & & & & \\
\hline	 
\end{tabular}
\end{table*}

\begin{figure*}
\centering
\begin{picture} (0,280) (260,0)
\includegraphics[width=1.0\textwidth]{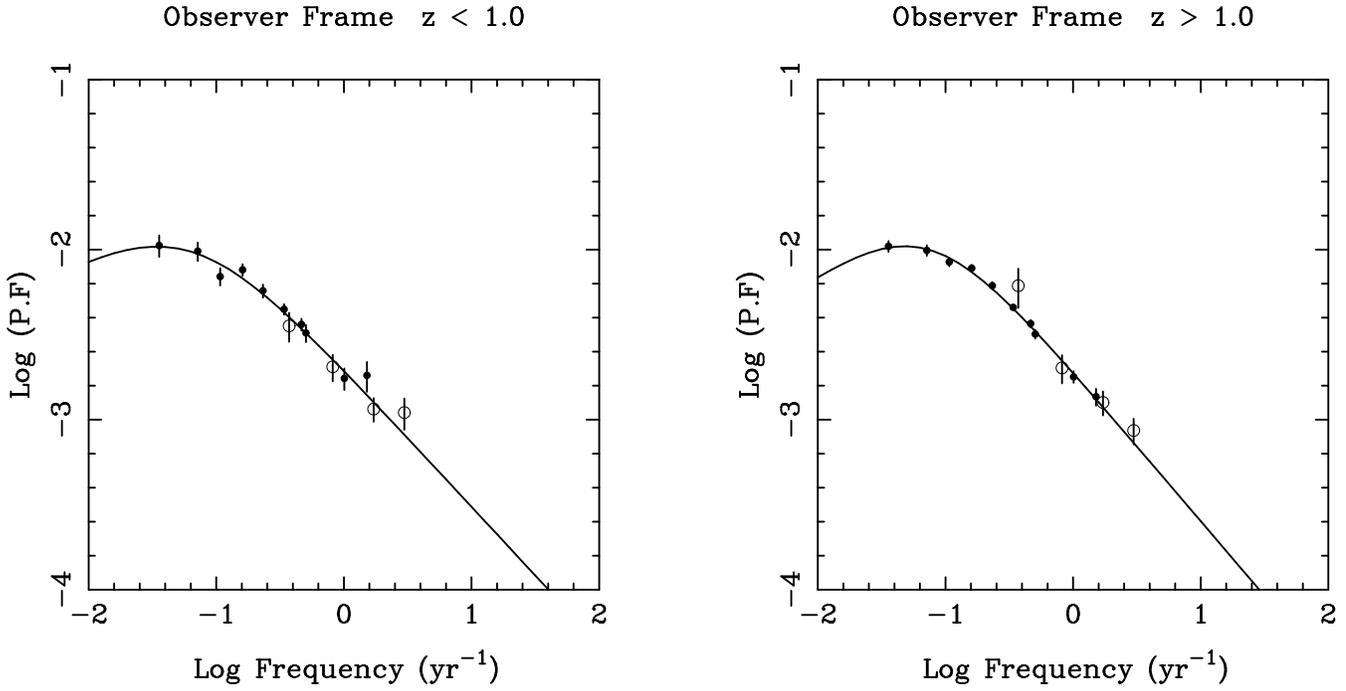}
\end{picture}
\caption{The left and right hand panel show SEDs of light curves in the
 observer frame for low and high redshift samples of quasars
 respectively. The symbols and solid line are as for the left hand
 panel of Fig.~\ref{fig1}.}
\label{fig3}
\end{figure*}

The SEDs in Fig.~\ref{fig1} represent timescales in the observer's
reference frame, and describe variations as measured from Earth.  If it
is assumed that these variations in flux are intrinsic to the quasars,
then the timescales will be subject to the effects of time dilation.
If we wish to recover the timescales as they would be seen in the rest
frame of the quasar, it is necessary to change the unit of time for
each individual light curve for a quasar of redshift $z$ by a factor
$(1+z)^{-1}$, and hence the frequency scale by a factor $(1+z)$.  This
leads to a modification of Eq.~\ref{eqn1} such that

\begin{equation}
\tau \rightarrow \frac{\tau}{1+z}
\label{eqn4}
\end{equation}

\noindent
As well as re-scaling by a factor of $(1+z)$ along the time axis, there
is an additional correction to make to the normalisation by a factor
of $(1+z)$, to keep the power the same in the observed and time-shifted
light curves.

Fig.~\ref{fig2} shows the effect of applying the correction for time
dilation to the data in Fig.~\ref{fig1}.  The SEDs are fitted with
the function P(f) defined in Eq.~\ref{eqn2} as before, and the
best-fit parameters are given in Table~\ref{tab1}.  Broadly speaking,
the effect of correcting for time dilation is to shift each of the
curves to shorter timescales by a factor $(1+<z>)$, where $<z>$ is
the mean redshift of the objects in the magnitude bins.  There is a
marked increase in the scatter in the data after they have been
corrected for time dilation, but the maximum power density is not
changed.  In order to quantify any shift in timescale after correcting
for time dilation, we define a reference time $t_{ref}$ as the 
timescale at which $log (P(f)) = -2.4$.  Values of $t_{ref}$ are also
given in Table~\ref{tab1}.

\subsection{Measurements of time dilation}

\begin{figure*}
\centering
\begin{picture} (0,280) (260,0)
\includegraphics[width=1.0\textwidth]{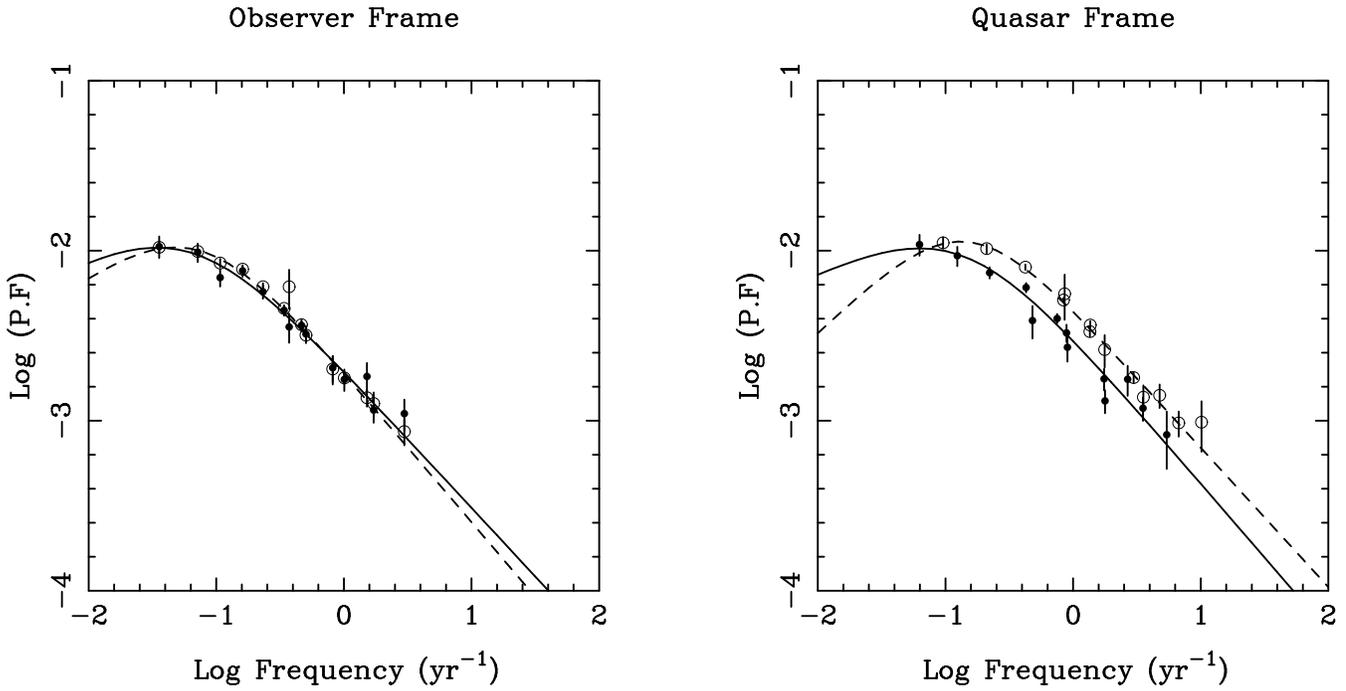}
\end{picture}
\caption{The left hand panel shows the superposition of the SEDs for
 low and high redshift samples of light curves from Fig.~\ref{fig1}.
 The right hand panel shows SEDs of the same light curves in the quasar
 frame, with a correction for time dilation applied.} 
\label{fig4}
\end{figure*}

One of the difficulties in looking for correlations in the properties
of quasars is the well known degeneracy between redshift and luminosity
in magnitude limited samples.  The degeneracy takes the form of a
correlation between redshift and luminosity such that in most available
samples the higher redshift members tend to be the more luminous.  Thus
if we are looking for a correlation between redshift and timescale, as
in time dilation, we must be careful that we do not actually measure a
correlation between luminosity and timescale.  We have shown in
Section~\ref{sec3} above that for the samples of quasars under
consideration here, we find an anti-correlation between quasar
luminosity and Fourier power density.  This in itself would not
necessarily be a problem, but the SEDs also show a correlation between
luminosity and the rate at which short timescale power density declines.
Given the correlation between redshift and luminosity, such an effect
could be confusing in a search for a correlation between redshift and
timescale as expected for time dilation.  To minimise any such effect
the absolute magnitude range of quasars was restricted to the band 
$-25.5 < M_{B} < -22.5$, which is sufficiently small that no
correlation of amplitude with luminosity is detectable.  This also has
the effect of reducing any contamination from host galaxy light to a
negligible level.

In order to measure the effects of time dilation we split the quasar
light curves into low and high redshift samples.  The idea was to
compare the resulting SEDs to look for the expected shift of the high
redshift sample towards longer timescales relative to the low redshift
sample.  Fig.~\ref{fig3} shows the low and high redshift SEDs
separately, and it can be seen that in spite of the restriction in
luminosity the SEDs are well defined, with excellent agreement between
the different datasets where they overlap.  The data are well fitted by 
the function $P(f)$ in Eq.~\ref{eqn2}, and the fit parameters of
interest are given in Table~\ref{tab1}.

Fig.~\ref{fig4} shows the effect of correcting the SEDs in
Fig.~\ref{fig3} for time dilation.  The left hand panel shows the two
SEDs in the observer frame superimposed to show any shift in timescale.
The right hand panel shows the same data with the correction for time
dilation applied.  It is immediately clear that in the observer frame
the two SEDs are very closely matched, with $t_{ref} = 2.2$ years in
both cases.  However, in the quasar frame there is a marked difference, 
the high redshift quasars being preferentially shifted to shorter
timescales.  It will be noticed that the maximum power density of both
high and low redshift SEDs is the same in the quasar and observer
frames, implying that the shift in the position of the SEDs is in
timescale.  The ratio of the values of $t_{ref}$ for the high and low
redshift samples is almost exactly equal to the ratio of the values for
$(1+<z>)$, implying that the effect of correcting for time dilation has
been to move the SEDs to shorter timescales by a factor of $(1+<z>)$,
with no other obvious change.

\section{Interpretation of results}
\label{sec5}

The results of Section~\ref{sec4} provide strong evidence that the
effects of time dilation are not seen in quasar light curves.  This
clearly runs against expectations based on a conventional cosmological
viewpoint, and so in this section we examine ways in which the results
may be understood.

\subsection{Black hole growth}

Perhaps the most straightforward way of explaining the absence of the
effects of time dilation in quasar light curves is to postulate an
increase in timescale of variation associated with the growth of the
central supermassive black hole of the AGN.  Thus higher redshift
quasars would contain less massive black holes which would vary more
quickly in such a way as to offset the effects of time dilation.  The
problem with this picture is that there is a well-supported correlation
between black hole mass and luminosity based on reverberation mapping
\cite{k00}.  This means that, given the restricted magnitude range of
our sample, there can be little difference in the average black hole
mass of the high and low redshift samples.  Even if we ignore the
restriction on luminosity, it would be difficult to cancel out time
dilation effects by assuming an increasing luminosity with redshift as
it is clear from Fig.~\ref{fig2} that the whole shape of the SED changes
with luminosity, especially the power law index to shorter timescales.
This is not what is seen in Fig.~\ref{fig4}, where the shape of the SEDs
does not change between high and low redshift samples.

\subsection{Microlensing}

Another possibility for explaining the absence of time dilation effects
in quasar light curves is that the variations do not predominantly
originate in the quasars, but along the line of sight at low redshift.
The most plausible mechanism for this is microlensing of the quasars by
a population of stellar mass bodies \cite{h07}, where the most probable
redshift for the lenses is $z \sim 0.5$ \cite{t84}.  Although such
microlensing is unambiguously seen in multiple images of gravitationally
lensed systems \cite{p98}, there are two main difficulties with
this approach.  Firstly, although the observed variations agree well with
model predictions from microlensing simulations, it is difficult to rule
out the possibility of intrinsic variations.  Secondly, it appears that
the rate of detection of compact bodies in the Galactic halo by the MACHO
project \cite{a97} is incompatible with the population required to
produce the observed variation in the quasar light curves. 

\subsection{Static universe}

The well known dilemma that Einstein was faced with when he realised that
his equations implied an expanding universe is still perhaps the best
reason for believing that the universe is not static, given the success
of general relativity in explaining cosmological observations.  There is
however surprisingly little direct evidence that the Universe is
expanding.  As mentioned in Section~\ref{sec1}, searches for time
dilation in gamma-ray bursts do not provide a conclusive test.  Supernova
light curves on the other hand appear to show convincing evidence of time
dilation \cite{f05}, which would rule out a non-expanding universe as
an explanation for the results presented here for quasar light curves.
Although this result has been challenged in an interesting paper by
\cite{c09} on the basis of bias in the sampling procedure, it seems fair
to say that the result is still generally accepted.  

\subsection{Quasar distances}

For completeness we should add that a possible explanation for the
apparent lack of time dilation effects in quasar light curves is that
quasars are not at the cosmological distances implied by their redshifts.
This idea has been energetically pursued by examining apparent
associations of quasars with relatively nearby galaxies or clusters
\cite{a01a}, but the large body of observations of quasar host galaxies
seems to rule out the possibility that quasars are nearby, and that as a
result time dilation would be negligible. 

\section{Discussion}
\label{sec6}

Taken at face value, the observations described above can only be
explained by at least a small departure from the conventional
cosmological view, and perhaps a large one.  It is therefore worth
reviewing the security of the observational results to be sure that they
are robust.  We need to check that the data are showing a consistent
picture.  We first note that the Field 287 and MACHO data are completely
independent of each other in that they contain different sets of
quasars selected according to different criteria in different areas of
sky.  The photometric monitoring was carried out using completely
different techniques with different error distributions, and yet for all
the SEDs presented above, in the area of overlap the agreement between
the two datasets is excellent.  The implication of this is that the shape
of the SEDs represents a true signal, and is not dominated by systematic
effects.

In the right hand panel of Fig.~\ref{fig1} the only difference between
the three SEDs is the quasar luminosity, and yet a clear progression in
slope of the power law index at short timescales is evident, implying a
real and measurable change in the nature of the variability, and not an
artefact of the analysis procedure.  This provides sound support for
believing that the close similarity of the high and low redshift SEDs in
Fig.~\ref{fig4} represents indistinguishable patterns of variability,
unaffected by any time dilation effect, and not a spurious agreement.

If it is accepted that time dilation is not seen in quasar light curves,
then some departure from conventional cosmology is necessary to explain
it.  Of the possibilities listed in Section~\ref{sec5}, there seems to be
overwhelming evidence that quasars are at the cosmological distances
indicated by their redshifts, and the challenge to the time dilation
found in supernovae light curves has yet to be convincingly established.
If we therefore assume that we live in an expanding universe, we have
two possibilities.  If the variations are due to microlensing then the
conclusions of the MACHO project would have to be modified, presumably by
a reassessment of the shape of the Galactic halo, and the expected dark
matter content.  If the effects of time dilation are offset by an
increase of timescale of variation with cosmological time, then a
mechanism must be found which does not alter the shape of the SED, or
involve a correlation of black hole mass with luminosity. 

\section{Conclusions}
\label{sec7}

In this paper we have used Fourier power spectrum analysis of over 800
quasar light curves to measure timescales of variation at different
redshifts.  The expected effects of time dilation are absent, the SEDs at
high and low redshift being essentially identical.  There seems to be no
explanation for this within the conventional cosmological framework, and
so various other possibilities are considered.  These include the idea
that the effects of time dilation are exactly offset by an increase
in timescale of variation associated with black hole growth.
Alternatively, the observed variations could be caused by microlensing,
in which case time dilation would not be expected.

\section*{Acknowledgements}

I thank Alan Heavens for suggesting the procedure for verifying the
effectiveness of the sampling correction.


\begin{thebibliography}{}

\bibitem[Alcock et al.\ 1997]{a97} Alcock C. et al., 1997, ApJ, 486,
 697

\bibitem[Alcock et al.\ 2001]{a01} Alcock C. et al., 2001, ApJ, 550,
 L169

\bibitem[Arp \& Russell 2001]{a01a} Arp H., \& Russell, D. 2001, ApJ,  549, 802

\bibitem[Chang 2001]{c01} Chang, H.-Y. 2001, ApJ, 557, L85

\bibitem[Crawford 2009]{c09} Crawford, D.F. 2009, arXiv:0901.4172

\bibitem[Cristiani et al.\ 1996]{c96} Cristiani, S., Trentini, S.,
 La Franca, F., Aretxaga, I., Andreani, P., Vio, R., Gemmo, A. 1996,
 A\&A, 306, 395

\bibitem[Deng \& Schaefer 1998]{d98} Deng, M., \& Schaefer, B.E.
 1998, ApJ, 502, L109

\bibitem[Dobrzycki et al.\ 2005]{d05} Dobrzycki, A., Eyer, L.,
 Stanek, K.Z., Macri, L.M. 2005, A\&A, 442, 495

\bibitem[Foley et al.\ 2005]{f05} Foley, R.J. et al. 2005, ApJ, 626, L11

\bibitem[Geha et al.\ 2003]{g03} Geha, M. et al. 2003, AJ, 125, 1

\bibitem[Goldhaber et al.\ 2001]{g01} Goldhaber, G. et al. 2001, ApJ,
 558, 359

\bibitem[Hawkins 2000]{h00} Hawkins, M.R.S. 2000, A\&AS, 143, 465

\bibitem[Hawkins 2007]{h07} Hawkins, M.R.S. 2007, A\&A, 462, 581

\bibitem[Hook et al.\ 1994]{h94} Hook, I.M., McMahon, R.G.,
 Boyle, B.J., Irwin, M.J. 1994, MNRAS, 268, 305
 
\bibitem[Kaspi et al.\ 2000]{k00} Kaspi, S., Smith, P.S., Netzer, H.,
 Maoz, D., Jannuzi, B.T., Giveon, U. 2000, ApJ, 533, 631

\bibitem[Pelt et al.\ 1998]{p98} Pelt, J., Schild, R., Refsdal, S.,
 Stabell, R. 1998, A\&A, 336, 829

\bibitem[Timmer \& K\"{o}nig 1995]{t95} Timmer, J., \& K\"{o}nig, M.
 1995, A\&A, 300, 707

\bibitem[Turner et al.\ 1984]{t84} Turner, E.L., Ostriker, J.P.,
 Gott, J.R. 1984, ApJ, 284, 1

\end{thebibliography}
\end{document}